\documentclass[twocolumn,showpacs,prb]{revtex4}
\usepackage{graphicx}%
\usepackage{dcolumn}
\usepackage{amsmath}
\usepackage{latexsym}
\begin{document}

\title{Komar energy and Smarr formula for noncommutative Schwarzschild black hole }

\author{Rabin Banerjee} 
\altaffiliation{rabin@bose.res.in}
\author{Sunandan Gangopadhyay}
\email{sunandan.gangopadhyay@gmail.com,sunandan@bose.res.in}
\affiliation{Satyendra Nath Bose National Centre for Basic Sciences, \\
Block-JD, Sector-III, Salt Lake, Kolkata - 700098, India;\\}
\affiliation{Department of Physics and Astrophysics, \\
West Bengal State University, Barasat, India\\
and Visiting Associate in Satyendra Nath Bose National Centre for Basic Sciences,\\
Block-JD, Sector-III, Salt Lake, Kolkata - 700098, India.}

%\altaffiliation{Visiting Associate in Satyendra Nath Bose National Centre for
%Basic Sciences, Kolkata, India}

\date{\today}

\begin{abstract}
\noindent
We calculate the Komar energy $E$ for a noncommutative
Schwarzschild black hole. A deformation from the conventional
identity $E=2ST_H$ is found in the next to leading order computation
in the noncommutative parameter $\theta$ 
(i.e. $\mathcal{O}(\sqrt{\theta}e^{-M^2/\theta})$) which is also consistent
with the fact that the area law now breaks down. This deformation yields
a nonvanishing Komar energy at the extremal point $T_{H}=0$ of these black holes.
We then work out the Smarr formula, clearly elaborating the differences from the standard
result $M=2ST_H$, where the mass ($M$) of the black hole is identified with the
asymptotic limit of the Komar energy. Similar conclusions are also shown to hold for a deSitter--Schwarzschild geometry.
\end{abstract}
\pacs{11.10.Nx}
%]
\maketitle

%%%%%%%%%%%%%%%%%%%%%%%%%%%%%%%%%%%%%%%%%%%%%%%%%%%%%%%%%%%%%%
\noindent The deep connection between gravity and thermodynamics  
has been known for a long time \cite{sak}--\cite{paddy}. 
For instance, in a thermodynamical system the entropy $S$,
temperature $T$ and energy $E$ are related by the first law
of thermodynamics
\begin{eqnarray}
dE &=& TdS
\label{intro1}
\end{eqnarray}
where, for simplicity, we ignore work terms.

\noindent A similar relation exists for black holes where $E$
is identified with the Komar energy \cite{komar},\cite{wald} and $T$ is the Hawking temperature.
Indeed there is an integral version of the above equation 
\begin{eqnarray}
E &=& 2TS~.
\label{intro2}
\end{eqnarray}
Specifically, for a Schwarzschild black hole, the Komar energy
equals the mass ($M$) of the black hole, so that
\begin{eqnarray}
M &=& 2TS~.
\label{intro3}
\end{eqnarray}
This identity is easily verified by putting the well known
expressions for the entropy and the Hawking temperature
\begin{eqnarray}
T&=&\frac{1}{8\pi M}\quad,\quad S=\frac{A}{4}=4\pi M^2 ~.
\label{intro4}
\end{eqnarray}
The relations (\ref{intro2}) and (\ref{intro3}) are quite fundamental
in black hole thermodynamics. The first one (\ref{intro2}) is very general
and holds for any black hole. General discussions and different derivations
of eq.(\ref{intro2}) have been provided earlier in the literature \cite{padma},\cite{rbstat}. 
The effective Komar energy appearing in the left hand side of eq.(\ref{intro2})
is basically identified with the conserved charge corresponding to the Killing vector defined at
the event horizon \cite{bmms}.
The relation (\ref{intro3}), on the contrary, is valid only for a Schwarzschild
black hole. It is known as the Smarr formula \cite{smarr} that connects the
macroscopic parameters of a black hole with $2TS$ (the right hand side of eq.(\ref{intro3})).
The connection between eq.(\ref{intro2}) and the generalised Smarr formula valid for any black hole
has been recently elaborated in a series of papers \cite{rbstat},\cite{bmms},\cite{ms}. 

In this paper we first present a short derivation of eq.(\ref{intro2}) 
for the specific case of a spherically symmetric metric,
where the role of the area law ($S=A/4$) is emphasised.
Then we consider the status  of eqs.(\ref{intro2}, \ref{intro3}) in the
context of noncommutative Schwarzschild black hole. It is well known that the
noncommutativity may be introduced in different ways \cite{spal}-\cite{rbreview}
in the context of black holes. Here we invoke noncommutativity through a coherent state formalism
\cite{sunfgs}, \cite{sunPLB} which implies the replacement of the ordinary
point wise multiplication by a Voros multiplication \cite{voros}. The analysis
is performed by carrying out an expansion in the noncommutative parameter $\theta$.
We show that both eqs.(\ref{intro2}, \ref{intro3}) are deformed. This deformation
starts from the next to leading order in the expansion process 
(i.e. from $\mathcal{O}(\sqrt{\theta}e^{-M^2/\theta})$) in case of relation (\ref{intro2}).
Indeed it is precisely at this order that the area law also breaks down. Consequently the implication
of the area law in the proof of eq.(\ref{intro2}), as shown here, gets highlighted.
Deformations in eq.(\ref{intro3}), on the contrary, are found from the leading order correction itself.
We find that the deviations from eqs.(\ref{intro2}, \ref{intro3}) begin to play
an important role at the extremal point ($T_{H}=0$) of these black holes.
We observe that as we go nearer to the extremal point (which corresponds to
a minimum value of the horizon radius $r_h$), the correction term 
involving the noncommutative parameter $\theta$ leads to a nonvanishing Komar
energy. 

The final part of this paper is dedicated to studying similar issues
in the context of deSitter--Schwarzschild geometry \cite{dymnikova}. The
reason is the similarity between this geometry with the noncommutative
Schwarzschild geometry. Both avoid the singularity problems, though
in different ways, present in the usual Schwarzschild black hole.
Also, as was shown in \cite{majhimodak}, the noncommutative Schwarzschild geometry
in the limit $r<<\sqrt{\theta}$ passes over to the deSitter metric. 
Our calculations show that for deSitter--Schwarzschild black holes \cite{dymnikova},
deviations from the expected relations (\ref{intro2}, \ref{intro3}) occur, quite analogous
to the noncommutative case. This is reassuring in the sense that it gives us confidence in the
construction of noncommutative black holes.

In \cite{sunPLB}, it has been shown that the expression for the Komar
energy $E$ of a spherically symmetric stationary black hole metric 
\begin{eqnarray}
ds^2 = -f(r) dt^2 + f^{-1}(r)dr^2 + 
r^2(d\tilde\theta^2+\sin^2\tilde\theta d\phi^2) 
\label{1.04}
\end{eqnarray}
is given by 
\begin{eqnarray}
E=\frac{1}{8\pi}\int_{\tilde\theta=0}^{\pi}
\int_{\phi=0}^{2\pi} r^2 \sin{\tilde\theta}~\partial_{r}f(r)
d\tilde\theta d\phi=\frac{r^2}{2}\partial_{r}f(r)~.
\label{komar}
\end{eqnarray}
Near the event horizon $r_h$ of the black hole, the above expression
for $E$ simplifies to
\begin{eqnarray}
E=r_{h}^2 \kappa=2\pi r_{h}^{2}T_{H}
\label{komar_hor}
\end{eqnarray}
where 
\begin{eqnarray}
\kappa=\frac{1}{2}[\partial_{r}f(r)]/_{r=r_{h}} 
\label{surf_grav}
\end{eqnarray}
is the surface gravity of the black hole and 
\begin{eqnarray}
T_{H}=\kappa/(2\pi)
\label{Hawk_temp}
\end{eqnarray}
is the Hawking temperature 
for a general static and spherically symmetric spacetime. 
Now assuming that the area law $S=A/4=\pi r_{h}^2$ holds,
it is easy to see that the
above relation for the Komar energy can be written as
\begin{eqnarray}
E=2ST_{H}
\label{komar_ener}
\end{eqnarray}
which is the sought relation.
 
We now proceed to investigate the status of 
the above relation in the case of noncommutative
black holes.
We start with the noncommutative Schwarzschild metric 
\cite{spal},\cite{sunPLB}  
\begin{eqnarray}
f(r)&=&f_{\theta}(r)=1-\frac{2M(\theta)}{r}\nonumber\\
M(\theta)&=&\frac{2M}{\sqrt\pi}\gamma\left(\frac{3}{2},\frac{r^2}{4\theta}\right)
\label{metric_coef}
\end{eqnarray}
where 
\begin{eqnarray}
\gamma(3/2, r^2/4\theta)=\int_{0}^{r^2/(4\theta)}dt~t^{1/2}e^{-t}
\label{incom_gamma}
\end{eqnarray}
is the lower incomplete gamma function.  
The Komar energy for this spherically symmetric spacetime can be computed from
eq.(\ref{komar}) and yields
\begin{eqnarray}
E&=&M(\theta)-\frac{Mr^3}{2\theta\sqrt{\pi\theta}}e^{-r^2/(4\theta)}
\label{komar_comp}
\end{eqnarray}
which in the limit $r\rightarrow\infty$ gives $E=M$. 
Also, expectedly, for $\theta\rightarrow0$, we reproduce $E=M$.
Now since the Komar
energy in the asymptotic limit ($r\rightarrow\infty$) gives the mass of the
black hole, therefore we identify $M$ (and not $M(\theta)$) as the mass of the
noncommutative Schwarzschild black hole (\ref{metric_coef}). This identication will play an important
role in the subsequent discussion.

\noindent The event horizon of the black hole can be 
found by setting $f_{\theta}(r_h)=0$ in eq.(\ref{metric_coef}), which yields
\begin{eqnarray}
r_h=\frac{4M}{\sqrt\pi}\gamma\left(\frac{3}{2},\frac{r^2_h}{4\theta}\right).
\label{1.05}
\end{eqnarray}
Since this equation cannot be solved in a closed form, we take 
the large radius regime ($\frac{r_h^{2}}{4\theta}>>1$) 
where we can expand the incomplete gamma function to 
solve $r_h$ by iteration. We  keep terms upto the leading 
($\frac{1}{\sqrt{\theta}}e^{-{M^2}/{\theta}}$) and next to leading
($\sqrt{\theta}e^{-{M^2}/{\theta}}$) orders, to obtain
\begin{eqnarray}
r_h \simeq 2M\left[1-\frac{2M}{\sqrt{\pi\theta}}
\left(1+\frac{\theta}{2M^{2}}\right)e^{{-M^2}/{\theta}}
\right]~. 
\label{1.06}
\end{eqnarray}
Using eqs.(\ref{surf_grav}),(\ref{Hawk_temp}) and (\ref{metric_coef}),
the Hawking temperature for the 
noncommutative Schwarzschild black hole is found to be
\begin{eqnarray}
T_H &=& {\frac{1}{4\pi}}\left[{\frac{1}{r_h}}-
{\frac{r_h^2}{4\theta^{3/2}}}\frac{e^-{\frac{{r_h}^2}{4\theta}}}
{\gamma({\frac{3}{2}},{\frac{r^2_h}{4\theta}})}\right].
\label{1.08}
\end{eqnarray}
To write the Hawking temperature in the regime 
$\frac{r^2_h}{4\theta}>>1$ as a function of $M$ 
we use (\ref{1.06}). Keeping, as before, terms upto the order 
$\sqrt{\theta}e^{-{M^2}/{\theta}}$, we get
\begin{eqnarray}
T_{H}=\frac{1}{8{\pi}M}
\left[1-\frac{4M^3}{{\theta}{\sqrt{\pi\theta}}}
\left(1-\frac{\theta}{2M^{2}}-\frac{\theta^2}{4M^{4}}\right)
{e^{-M^2/\theta}}\right]~.\nonumber\\
\label{1.10}
\end{eqnarray}
We shall now use the first law of black hole thermodynamics 
to calculate the Bekenstein-Hawking entropy. 
This law is given by \cite{Beken3} 
\begin{eqnarray}
dS=\frac{dM}{T_H}~.
\label{1.1}
\end{eqnarray}
Hence the Bekenstein-Hawking entropy upto  
the order $\sqrt{\theta}e^{-{M^2}/{\theta}}$ is found to be 
\begin{eqnarray}
S=\int{\frac{dM}{T_H}}=4\pi M^2-16\sqrt{\frac{\pi}{\theta}}M^3
\left(1+\frac{\theta}{M^2}\right)e^{-M^2/\theta}~.\nonumber\\
\label{1.11}
\end{eqnarray}
In order to express the entropy in terms of the 
noncommutative horizon area ($A_{\theta}$), 
first we use eq.(\ref{1.06}) to obtain
\begin{eqnarray}
A_{\theta}&=& 4\pi r^2_h=16\pi M^2-64\sqrt{\frac{\pi}{\theta}}
M^3\left(1+\frac{\theta}{2M^2}\right)e^{-M^{2}/\theta}\nonumber\\
&&+\mathcal{O}(\theta^{3/2}e^{-M^{2}/\theta}).
\label{1.12}
\end{eqnarray}
Comparing eqs.(\ref{1.11}) and (\ref{1.12}), 
we find that at the leading order in $\theta$ 
(i.e. $\mathcal{O}(\frac{1}{\sqrt{\theta}}e^{-{M^2}/{\theta}})$), 
the noncommutative black hole entropy satisfies the area law
\begin{eqnarray}
S=\frac{A_{\theta}}{4}~.
\label{1.13}
\end{eqnarray}
However, it is easy to note that there is a deviation 
from the area law in the next to leading order in $\theta$, i.e. 
at $\mathcal{O}(\sqrt{\theta}e^{-{M^2}/{\theta}})$.

\noindent Now we compute the Komar energy at the horizon using 
eqs.(\ref{komar_hor}, \ref{1.06}, \ref{1.10}) upto the        
the order $\sqrt{\theta}e^{-{M^2}/{\theta}}$, to get
\begin{eqnarray}
E &=& M\left[1-\frac{2M}{\sqrt{\pi\theta}}
\left(\frac{2M^{2}}{\theta}+1\right)e^{{-M^2}/{\theta}}\right.\nonumber\\
&&\left.-\frac{1}{M}\sqrt{\frac{\theta}{\pi}}e^{{-M^2}/{\theta}}\right].\label{komar1c}
\end{eqnarray}
Finally, using eqs.(\ref{1.10}), (\ref{1.11}) and (\ref{komar1c}),       
we obtain
\begin{eqnarray}
E&=&2ST_{H} +2\sqrt{\frac{\theta}{\pi}}e^{{-M^2}/{\theta}}+ 
\mathcal{O}(\theta^{3/2}e^{-M^{2}/\theta})\nonumber\\
&=&2ST_{H} +2\sqrt{\frac{\theta}{\pi}}e^{{-S}/{(4\pi\theta)}}+ 
\mathcal{O}(\theta^{3/2}e^{-S/(4\pi\theta)})\nonumber\\
\label{komar2a}
\end{eqnarray}
where in the second line we have used eq.(\ref{1.11}) to replace $M^2$
by $S/(4\pi)$ in the exponent. 

The above relation can also be written 
with $M$ being expressed in terms of the black hole parameters
$S$ and $T_H$ using eq.(\ref{komar1c}) 
\begin{eqnarray}
M&=&2ST_{H} +\frac{1}{2\pi\sqrt{\pi\theta}}\left(S+\frac{S^2}{2\pi\theta}+6\pi\theta\right)e^{-S/(4\pi\theta)}\nonumber\\
&&+\mathcal{O}(\theta^{3/2}e^{-S/(4\pi\theta)}).
\label{nc_smarr}
\end{eqnarray} 
Since $M$ has been identified
earlier to be the mass of the black hole, we name eq.(\ref{nc_smarr}) 
as the {\it{Smarr formula for noncommutative Schwarzschild black hole}}. 
The above relations (\ref{komar2a}, \ref{nc_smarr}) are the 
main results of this paper.

\noindent We now make some observations on eq.(\ref{komar2a}). Interestingly, we have once again
been able to write the deformed relation (involving the
noncommutative parameter $\theta$) in terms of the Komar
energy $E$, entropy $S$ and the Hawking temperature $T_H$.
The above result  (\ref{komar2a}) shows that the 
relation (\ref{komar_ener}) is satisfied upto the leading order
in $\theta$ (i.e. $\mathcal{O}(\frac{1}{\sqrt{\theta}}e^{-{M^2}/{\theta}})$). 
This is indeed consistent with the fact that the
area law (\ref{1.13}) also holds at this order in $\theta$. However,
the validity of the relation (\ref{komar_ener}) breaks down in the
next to leading order in $\theta$ 
(i.e. $\mathcal{O}(\sqrt{\theta}e^{-{M^2}/{\theta}})$) 
in accordance with the fact that there is a deviation 
from the area law at this order in $\theta$
as seen above. The noncommutative Smarr formula (\ref{nc_smarr}), on the
contrary, deviates from the usual one (\ref{intro3}) right from the leading
correction itself.

\noindent Further, we note that for a black hole with an entropy
$S>>4\pi\theta$, the second term on the right hand side of
eq.(\ref{komar2a}) is small compared to the first term and hence the 
relation (\ref{komar_ener}) is once again satisfied. This is in
conformity with an observation made in \cite{majhimodak} where it
was shown that the area law (\ref{1.13}) is satisfied for
$r_{h}\geq4.8\sqrt{\theta}$ implying (from our present analysis) 
that the relation (\ref{komar_ener})
holds. To see this, we rewrite eq.(\ref{komar2a}) by replacing 
$S$ by $A_{\theta}/4$ in the exponent, to get
\begin{eqnarray}
E&=&2ST_{H} +2\sqrt{\frac{\theta}{\pi}}e^{{-A_\theta}/{(16\pi\theta)}}
+\mathcal{O}(\theta^{3/2}e^{-A_\theta/(16\pi\theta)})~.\nonumber\\
\label{komar2aq}
\end{eqnarray}
Therefore, when $A_{\theta}>>16\pi\theta$, i.e. $r_{h}>>2\sqrt{\theta}$,
the second term on the right hand side of
eq.(\ref{komar2aq}) is small compared to the first term and hence the 
relation (\ref{komar_ener}) gets satisfied which implies that
the area law (\ref{1.13}) holds in this regime.

\noindent Finally, note that there is a lower bound to the
horizon area $A_\theta$ ($A_\theta=k\theta$, $k\geq1$) from purely
physical grounds. This is also compatible with the fact that
there is a lower bound to the horizon radius $r_h$
($r_h\geq3\sqrt{\theta}$) since the Hawking temperature becomes negative
for $r_{h}<3\sqrt{\theta}$ \cite{majhimodak}. This yields $k\geq36\pi$.
Further, for $r_h =3\sqrt{\theta}$, the contribution to the Komar
energy comes from the second term in eq.(\ref{komar2aq}) only since $T_H =0$.
Putting the minimum value of $k$ (corresponding to which $T_H =0$)
in eq.(\ref{komar2aq}), we obtain
\begin{eqnarray}
E&=&2\sqrt{\frac{\theta}{\pi}}e^{-9/4}~.
\label{komar2aq1}
\end{eqnarray}
Interestingly, the above expression gives a nonvanishing Komar energy for
noncommutaive Schwarzschild black holes at their extremal point ($T_H =0$) which 
vanishes in the limit $\theta\rightarrow0$. The nonvanishing Komar energy thus owes
its origin to the noncommutative parameter $\theta$ appearing in the metric of the
noncommutative Schwarzschild black hole (\ref{metric_coef}).

\noindent To get further insight in the results obtained so far, 
we now consider the metric of deSitter-Schwarzschild geometry 
(given by Dymnikova \cite{dymnikova}) having similar features as the noncommutative
Schwarzschild metric (\ref{metric_coef}) at $r\rightarrow\infty$ and $r\rightarrow0$ limits. The
metric coefficient of this spherically symmeric black hole reads
%\begin{eqnarray}
%ds^2 &=& g(r)dt^2-g^{-1}(r)dr^2-r^2d\tilde\theta^2-r^2\sin^{2}\tilde\theta d\phi^2\nonumber\\
%\label{dymnikova}
%\end{eqnarray}
%where,
\begin{eqnarray}
\label{dym}
g(r)&=&1-\frac{r_g}{r}(1-e^{-r^3/(r_{0}^{2}r_g)})\\
r_g &=& 2M~,~ r_{0}^2 = 3/\Lambda.\nonumber
\end{eqnarray}
The above metric has the property that at $r\rightarrow\infty$, it behaves like Schwarzschild geometry
and at $r\rightarrow0$, it behaves like deSitter geometry and hence there is no singularity at $r=0$.
These properties are similar to the metric (\ref{metric_coef}) of the noncommutative Schwarzschild
black hole.

Once again the Komar energy for this spherically symmetric spacetime (\ref{dym}) can be computed from
eq.(\ref{komar}) and yields
\begin{eqnarray}
E&=&\frac{r_g}{2}(1-e^{-r^3/(r_{0}^{2}r_g)})-\frac{3r^3}{2r_{0}^2}e^{-r^3/(r_{0}^{2}r_g)}
\label{komar_comp_dym}
\end{eqnarray}
which in the limit $r\rightarrow\infty$ gives $E=M$. 
Therfore we identify $M$ as the mass of this black hole. 

\noindent The event horizon of the black hole can be 
found by setting $g(r^{(D)}_h)=0$\footnote{The superscript $D$ stands for Dymnikova.}, which leads to
\begin{eqnarray}
r^{(D)}_{h}&=&r_{g}(1-e^{-r^{(D)3}_h/(r_{0}^{2}r_g)})~.
\label{hor_raddym}
\end{eqnarray}
This equation once again cannot be solved for $r^{(D)}_h$ in a closed form. 
However, in the large radius regime $(r^{(D)3}_h/(r_{0}^{2}r_g)>>1)$, we can solve 
(\ref{hor_raddym}) by iteration. Keeping terms upto the order 
$e^{-r_{g}^2/r_{0}^{2}}=e^{-4M^2/r_{0}^{2}}$, we find
\begin{eqnarray}
r^{(D)}_{h}&=&r_{g}(1-e^{-r^{2}_g/r_{0}^{2}})~.
\label{hor_rad_100}
\end{eqnarray}
Using eqs.(\ref{surf_grav}, \ref{Hawk_temp}, \ref{dym}, \ref{hor_raddym}), the Hawking temperature for the 
deSitter-Schwarzschild black hole is found to be
\begin{eqnarray}
T^{(D)}_{H}&=&\frac{1}{4\pi}\left[\frac{r_g}{r^{(D)2}_{h}}(1-e^{-r^{(D)3}_h/(r_{0}^{2}r_g)})\right.\nonumber\\
&&\left.-\frac{3r^{(D)}_h}{r_{0}^2}e^{-r^{(D)3}_h/(r_{0}^{2}r_g)}\right]~.
\label{temp}
\end{eqnarray}
Keeping, as before, terms upto order $e^{-r_{g}^2/r_{0}^2}$, we get
\begin{eqnarray}
T^{(D)}_{H}&=&\frac{1}{4\pi r_g}\left[1 + e^{-r_{g}^2/r_{0}^{2}}
-\frac{3r_{g}^2}{r_{0}^2}e^{-r_{g}^2/r_{0}^{2}}\right]~.
\label{temp_approx}
\end{eqnarray}
The first law of black hole thermodynamics (\ref{1.1})
is now used to compute the entropy which upto the order $e^{-4M^2/r_{0}^2}$ yields  
\begin{eqnarray}
S^{(D)}=4\pi M^2-2\pi r_{0}^{2}e^{-4M^2/r_{0}^2}-12\pi M^{2}e^{-4M^2/r_{0}^2}~.\nonumber\\
\label{entropy_dym}
\end{eqnarray}
In order to express the entropy in terms of the 
horizon area ($A^{(D)}$), 
we use (\ref{hor_rad_100}) to obtain
\begin{eqnarray}
A^{(D)}=16\pi M^2(1-2\pi r_{0}^{2}e^{-4M^2/r_{0}^2}-2\pi M^{2}e^{-4M^2/r_{0}^2})~.\nonumber\\
\label{area_dym}
\end{eqnarray}
Comparing equations (\ref{entropy_dym}) and (\ref{area_dym}), 
we find that the area law is not satisfied at the order $e^{-4M^2/r_{0}^2}$.
As in the case of the noncommutative Schwarzschild black hole,
there is a deviation in this case from the area law 
at the order $e^{-4M^2/r_{0}^2}$.

\noindent Now to obtain the relation between the Komar energy $E$,
entropy $S$ and the Hawking temperature $T_H$,
we compute the Komar energy at the horizon using 
eqs.(\ref{komar_hor}, \ref{hor_rad_100}, \ref{temp}) upto the        
the order $e^{-4M^2/r_{0}^2}$, to get
\begin{eqnarray}
E^{(D)} &=&M\left(1-e^{-4M^2/r_{0}^2}-\frac{12M^2}{r_{0}^2}e^{-4M^2/r_{0}^2}\right).
\label{komar1cxzz}
\end{eqnarray}
Finally, using eqs.(\ref{temp}), \ref{entropy_dym}, \ref{komar1cxzz})       
and replacing $M^2$ by $S/(4\pi)$, we obtain
\begin{eqnarray}
E^{(D)}&=&2S^{(D)}T_{H}^{(D)} +\sqrt{\frac{S^{(D)}}{4\pi}}
\left(1+\frac{2\pi r_{0}^2}{S^{(D)}}\right)e^{-S^{(D)}/(\pi r_{0}^2)}\nonumber\\
&&+\mathcal{O}(e^{-2S^{(D)}/(\pi r_{0}^2})).
\label{komardym}
\end{eqnarray}
The above relation can also be written 
with $M$ being expressed in terms of the black hole parameters
$S^{(D)}$ and $T_{H}^{(D)}$ using eqs.(\ref{komar1cxzz}, \ref{entropy_dym}) 
\begin{eqnarray}
M&=&2ST_{H}+\sqrt{\frac{S^{(D)}}{4\pi}}
\left(2+\frac{2\pi r_{0}^2}{S^{(D)}}+\frac{12S^{(D)}}{4\pi r_{0}^2}\right)e^{-S^{(D)}/(\pi r_{0}^2)}\nonumber\\
&&+\mathcal{O}(e^{-2S^{(D)}}/(\pi r_{0}^2)).
\label{smarrdym}
\end{eqnarray} 
The above relations (\ref{komardym}, \ref{smarrdym}) are the analogues of eqs.(\ref{komar2a}, \ref{nc_smarr}). 
Note that (as in the case of the noncommutative Schwarzschild
black hole (\ref{komar2a})) once again we obtain a deformed relation involving the Komar
energy $E$, entropy $S$ and the Hawking temperature $T_H$ at the order $e^{-4M^2/r_{0}^2}$. 
This is consistent with the fact that the area law in this case is not satisfied at this order.

\noindent Now from eq.(\ref{temp}), we find that the Hawking temperature $T^{(D)}_H$ 
vanishes when $r^{(D)}_{h}=0$, $\infty$ for any $r_g$. Considering the physically
interesting solution $r^{(D)}_{h}=0$ (for which $T^{(D)}_H =0$), which also satisfies 
(\ref{hor_raddym}) for any $r_g$, we find that the 
relation (\ref{komardym}) upto order $\mathcal{O}(e^{-S^{(D)}/(\pi r_{0}^2)})$ reduces to
\begin{eqnarray}
E^{(D)}=\sqrt{\frac{S^{(D)}}{4\pi}}
\left(1+\frac{2\pi r_{0}^2}{S^{(D)}}\right)e^{-S^{(D)}/(\pi r_{0}^2)}.
\label{komar2aqjg}
\end{eqnarray}
Hence we find that there exists a nonvanishing Komar energy in case of deSitter-Schwarzschild
black holes when the Hawking temperature $T^{(D)}_H =0$.
This feature is exactly similar with the results obtained in
case of noncommutative Schwarzschild black holes. 
However, in the latter case, there is a nonvanishing
lower bound of the horizon radius at which the Hawking temperature vanishes whereas
the horizon radius goes to zero when the Hawking temperature vanishes in
the former case.

To summarise, we have computed the Komar energy for a noncommutative
Schwarzschild black hole governed by the metric (\ref{metric_coef}). This is used
to obtain (\ref{komar2a}) where a deformation from the standard result (\ref{intro2})
is found in the next to leading order in the expansion involving the noncommutative
parameter $\theta$. Next, the mass of the black hole is identified by taking the
asymptotic infinity limit of the Komar energy. This is found to be identical
with the usual Schwarzschild mass $M$. The noncommutative version of the Smarr
formula (\ref{nc_smarr}) is obtained. Here the deformation from the usual formula
(\ref{intro3}) begins at the leading noncommutative correction. Similar results are
derived for a deSitter-Schwarzschild geometry governed by the metric (\ref{dym}).
Since the noncommutative Schwarzschild metric, in an appropriate limit, passes over to this 
metric, such a similarity bolsters our confidence in carrying out computations in a noncommutative
formulation. Finally, we conclude by noting that the presence of a nonvanishing Komar
energy when the Hawking temperature vanishes in case of noncommutative Schwarzschild
black holes can also be found in the deSitter-Schwarzschild
geometry \cite{dymnikova} thereby concretizing our results.
Further, in the former case, the result is due to a combined effect
of the noncommutative parameter $\theta$ and the entropy of the black hole whereas the
result in case of deSitter-Schwarzschild metric is due to the cosmological constant and the 
entropy of the black hole.

%\vskip 0.5cm
%\noindent {\it{Acknowledgements}} : SG would like to
%thank the S.N. Bose National Centre for Basic Sciences, Kolkata, India
%for providing computer and library facilities during the stay as a Visiting Associate. 


\begin{thebibliography}{99}
\bibitem{sak}A.D. Sakharov, Sov. Phys. Dokl. 12, 1040 (1968).
\bibitem{Beken3} J.D.Bekenstein, Phys. Rev. D 7, 2333 (1973). 
\bibitem{Hawking1} S.W.Hawking, Nature 248, 30 (1974).
\bibitem{Hawking2} S.W.Hawking, Commun. Math. Phys. 43, 199 (1975).
\bibitem{Bardeen}J.M. Bardeen, B. Carter, S.W. Hawking, 
Commun. Math. Phys. 31, 161 (1973).
\bibitem{jacob}T. Jacobson, Phys. Rev. Lett. 75, 1260 (1995).
\bibitem{vol}G.E. Volovik, Phys. Rep. 351, 195 (2001).
\bibitem{visser}M. Visser, Mod. Phys. Lett. A 17, 977 (2002). 
\bibitem{pedro}E. Elizalde, J.S. Pedro, Phys. Rev. D 78, 061501 (2008).
\bibitem{hogan}C.J. Hogan, Phys. Rev. D 77, 104031 (2008).
\bibitem{paddy}D. Kothawala, T. Padmanabhan, S. Sarkar,
Phys. Rev. D 78, 104018 (2008).
\bibitem{komar}A. Komar, Phys. Rev. 113, 934 (1959).
\bibitem{wald}R.M. Wald, ``General Relativity", Chicago, U.S.A : Univ. Pr. (1984).
\bibitem{padma}T. Padmanabhan, Class. Quant. Grav. 21, 4485 (2004); [gr-qc/0308070].
\bibitem{rbstat}R. Banerjee, B.R. Majhi, ``Statistical Origin of Gravity",
[arXiv:1003.2312 [gr-qc]].
\bibitem{bmms}R. Banerjee, B.R. Majhi, S.K. Modak, S. Samanta, 
``Smarr Formula and Killing Symmetries for Black Holes in Arbitrary Dimensions", [arXiv: 1007.5204 [gr-qc]]. 
\bibitem{smarr} L. Smarr, Phys. Rev. Lett. 30, 71 (1973), [Erratum-ibid. 30, 521 (1973)].
\bibitem{ms}S.K. Modak, S. Samanta, ``Timelike Geodesic, Effective Komar Conserved Quantities and Entropy in Kerr-Newman black Hole",
[arXiv: 1006.3445 [gr-qc]].
\bibitem{spal} P.Nicolini, A.Smailagic, E.Spallucci, 
Phys. Lett. B 632, 547 (2006); [gr-qc/0510112].
\bibitem{spal2}S. Ansoldi, P. Nicolini, A. Smailagic, E. Spallucci,
Phys. Lett. B 645, 261 (2007).
\bibitem{chaichian} M. Chaichian, A. Tureanu, G. Zet, Phys. Lett. B 660: 573, (2008); [arXiv:0710.2075 [hep-th]].
\bibitem{anirban}P. Mukherjee, A. Saha, Phys. Rev. D 77: 064014, (2008); [arXiv:0710.5847 [hep-th]]. 
\bibitem{spalreview} P.Nicolini, Int. J. Mod. Phys A 24, 7 (2009) 1229; arXiv: 0807.1939 [hep-th].
\bibitem{samanta}R. Banerjee, B.R. Majhi, S. Samanta,
Phys. Rev. D 77, 124035 (2008), [arXiv:0801.3583 [hep-th]].
\bibitem{majhimodak}R. Banerjee, B.R. Majhi, S.K. Modak, 
Class. Quant. Grav., 26, 085010 (2009), [arXiv:0802.2176].
\bibitem{rbreview} R. Banerjee, B. Chakraborty, S. Ghosh, P. Mukherjee, S. Samanta, Found. Phys. 39: 1297, (2009);
e-Print: arXiv:0909.1000 [hep-th]. 
\bibitem{sunfgs}S. Gangopadhyay, F.G. Scholtz, Phys. Rev. Lett. 102, 241602 (2009); 
arXiv:0904.0379 [hep-th].
\bibitem{sunPLB}R. Banerjee, S. Gangopadhyay, S.K. Modak,
Phys. Lett. B 686, 181 (2010); [arXiv:0911.2123 [hep-th]].
\bibitem{voros}A. Voros, Phys. Rev. A 40 (1989) 6814.
\bibitem{dymnikova}I. Dymnikova, Gen. Rel. Grav. 24 (1992) 235.






%\bibitem{Beken1} J.D.Bekenstein, {\it{Ph.D. Thesis}} 
%Princeton University, Princeton, NJ (1972).
%\bibitem{Beken2} J.D.Bekenstein, 
%{\it{Lett. Nuovo Cimento}} {\bf{4}} 737 (1972).

%\bibitem{Beken4} J.D.Bekenstein, {\it{Phys. Rev.}} {\bf{D 9}} 3292 (1974).
%\bibitem{Modakex}R. Banerjee, S.K. Modak, {\it JHEP} {\bf{0905}}, 063 (2009), [arXiv:0903.3321].
%\bibitem{kaul} R.K. Kaul, P. Majumdar, {\it{Phys. Rev. Lett.}} {\bf{84}} 5255 (2000); [gr-qc/0002040].
%\bibitem{ghoshmitra}A. Ghosh, P. Mitra, {\it{Phys. Lett.}} {\bf{B 616}} 114 (2005); [gr-qc/0411035].
%\bibitem{list}For a review and complete list of papers on logarithmic corrections, see D.N. Page, {\it{New J. Phys.}} {\bf{7}} 203 (2005);
%[hep-th/0409024].
%\bibitem{Lousto}C.O. Lousto, Norma G. Sanchez, {\it Phys.Lett.} {\bf B212}, 411 (1988). 
%
%\bibitem{Smailrev}P.Nicolini, {\it{Int. J. Mod. Phys.}} {\bf{A24}}: 1229, 2009; arXiv:0807.1939 [hep-th].
%\bibitem{Banrev}R. Banerjee, B. Chakraborty, S. Ghosh, P. Mukherjee, S. Samanta {\it Found.Phys.} {\bf 39}, 1297 (2009), [arXiv:0909.1000 [hep-th]]. 
%%\bibitem{Voros} A. Voros, {\it Phys. Rev.} {\bf A 40}, 6814 (1989).
%\bibitem{Lizzi}S. Galluccio, F. Lizzi, P. Vitale, {\it{Phys. Rev. }} 
%{\bf {D 78}}: 085007 (2008).
%\bibitem{gouba}F.G. Scholtz, L. Gouba, A. Hafver, C.M. Rohwer, 
%{\it J. Phys.} {\bf A 42} 175303 (2009).
%\bibitem{sunandan}S. Gangopadhyay, F.G. Scholtz, {\it Phys. Rev. Lett.} {\bf 102}, 241602 (2009). 
%\bibitem{Fursaev}D.V.Fursaev, {\it Phys. Rev.} {\bf D 51}, R5352 (1995) [arXiv:hep-th/9412161].\\
 %                R.B.Mann and S.N.Solodukhin, {\it Nucl. Phys.} {\bf B 523}, 293 (1998) [arXiv:hep-th/9709064].
%\bibitem{Partha}R.K.Kaul and P.Majumdar, {\it Phys. Rev. Lett.} {\bf 84}, 5255 (2000) [arXiv:gr-qc/0002040].\\
 %               K.A Meissner, {\it Class. Quantum. Grav.}{\bf 21}, 5245 (2004) [arXiv:gr-qc/0407052].
%\bibitem{Das} Saurya Das, Parthasarathi Majumdar, Rajat K. Bhaduri, {\it Class. Quant. Grav.} {\bf 19}, 2355  (2002) [arXiv:hep-th/0111001].\\
 %             S.S.More, {\it Class. Quantum Grav.} {\bf 22}, 4129 (2005) [gr-qc/0410071].\\
  %            S.Mukherjee and S.S.Pal, {\it JHEP} {\bf 0205}, 026 (2002) [arXiv:hep-th/0205164]. 
%\bibitem{Carlip}S Carlip, {\it Class. Quantum Grav.} {\bf 17}, 4175 (2000) [arXiv: gr-qc/0005017].\\
 %               M.R. Setare, {\it Eur.Phys.J.C} {\bf 33}, 2004 [arXiv:hep-th/0309134] 
%\bibitem{Hooft}G. ’t Hooft, {\it Int. J. Mod. Phys.} {\bf A} 11, 4623 (1996), [arXiv:gr-qc/9607022].\\
 %              S. Sarkar, S. Shankaranarayanan, L. Sriramkumar, {\it Phys.Rev.} {\bf D} 78, 024003 (2008), arXiv:0710.2013 [gr-qc]. 
%\bibitem {Hawking3}S.W.Hawking, {\it Commun. Math. Phys.} {\bf 55}, 133 (1977). 
%\bibitem{Dewitt} Bryce S. Dewitt, {\it Phys. Rep.} {\bf 19}, 295 (1975).
%\bibitem{Parikh}M.~Parikh, F.~Wilczek,{\it{Phys.\ Rev.\ Lett.}} {\bf{85}}, 5042 (2000).
%\bibitem{tunneling}E.~C.~Vagenas, {\it{Phys.\ Lett.}} {\bf{B 503}}, 399 (2001); {\it{Mod.\ Phys.\ Lett.}} {\bf{A 17}}, 609 (2002); {\it{Phys.\ Lett.}} {\bf{B 533}}, 302 (2002).
%\bibitem{Majhibeyond}R. Banerjee, B.R. Majhi, {\it JHEP} {\bf{0806}}, 095 (2008), [arXiv:0805.2220].
%\bibitem{Modakbtz}S.K. Modak, {\it Phys. Lett.} {\bf B 671}, 167 (2009), [arXiv:0807.0959].
%\bibitem{Majhifermion}B.R. Majhi, {\it Phys.Rev.}{\bf D 79}, 044005 (2009), [arXiv:0809.1508]
%\bibitem{Majhitrace}R. Banerjee, B.R. Majhi, {\it Phys. Lett.} {\bf B} 674, 218 (2009), [arXiv:0808.3688].
%\bibitem{banmajhi}R. Banerjee, B.R. Majhi, {\it{Phys. Lett.}} {\bf{B 662}} 62 (2008); [arXiv:0801.0200].
%\bibitem{Modaklovelock}R. Banerjee, S.K. Modak, {\it JHEP} 0911, 073 (2009), [arXiv:0908.2346].
%\bibitem{Tao}T. Zhu, Ji-R. Ren, Ming-F. Li, 
%{\it JCAP} 0908, 010 (2009) [arXiv:0905.1838].
%\bibitem{gamboa}J. Gamboa, M. Loewe, J.C. Rojas, {\it{Phys. Rev.}}
%{\bf{D 64}} 067901, 2001.
%\bibitem{riv}V. Rivelles, {\it{Phys.\ Lett.}} {\bf{B 558}} 191 (2003).
%\bibitem{wess1}P. Aschieri, C. Blohman, M. Dimitrijeric, F. Meyer, P. Schupp,
%J. Wess, {\it{Class. Quant. Grav.}} {\bf{22}}: 3511 (2005).
%\bibitem{wess2}P. Aschieri, M. Dimitrijeric, F. Meyer, S. Schraml
%J. Wess, {\it{Lett. Math. Phys.}} {\bf{78}}: 61 (2006).
%\bibitem{lopez}J.C. Lopez-Dominguez, O. Obregon, M. Sabido, C. Ramirez, {\it{Phys. Rev.}} {\bf{D 74}}, 084024 (2006); [hep-th/0607002].
%\bibitem{okon}C. Chryssomalakos, E. Okon, {\it{JHEP}} {\bf{0708}}: 012, 2007.
%\bibitem{lopez}J.C. Lopez-Dominguez, O. Obregon, M. Sabido, C. Ramirez, {\it{Phys. Rev.}} {\bf{D 74}}, 084024 (2006); [hep-th/0607002].
%\bibitem{chaichian}M. Chaichian, A. Tureanu, G. Zet, {\it{Phys.\ Lett.}} {\bf{B 660}} 573 (2008); [arXiv:0710.2075].
%\bibitem{saha}P. Mukherjee, A. Saha, {\it{Phys. Rev.}} {\bf{D 77}} 064014; [arXiv:0710.5847].
%\bibitem{muk}R. Banerjee, P. Mukherjee, S. Samanta, {\it{Phys. Rev.}} {\bf{D 75}} 125020 (2007); [hep-th/0703128].
%\bibitem{szabo}R.J. Szabo, {\it{Phys. Rep.}} 
%{\bf{378}}, 207 (2003);[hep-th/0109162].
%\bibitem{sun1}B. Chakraborty, S. Gangopadhyay, A. Saha, 
%{\it{Phys. Rev.}} {\bf{D 70}} 107707 (2004); [hep-th/0312292];
%F.G. Scholtz, B. Chakraborty, S. Gangopadhyay, A.G. Hazra, 
%{\it{Phys. Rev.}} {\bf{D 71}} 085005 (2005); [hep-th/0502143];
%S. Gangopadhyay, ``Some Studies in Noncommutative Quantum Field Theories
%", (Ph.D. Thesis) arXiv:0806.2013[hep-th].
%\bibitem{chaichian1}M. Chaichian, P. Kulish, K. Nishijima,
%A. Tureanu, {\it{Phys. Lett.}} {\bf{B 604}} 
%98 (2004); [hep-th/0408069].
%\bibitem{chaichian2}M. Chaichian, P. Prenajder, A. Tureanu, 
%{\it{Phys. Rev. Lett.}} {\bf{94}} 
%151602 (2005); [hep-th/0409096].
%\bibitem{chaichian3}M. Chaichian, A. Tureanu, {\it{Phys. Lett.}} {\bf{B 637}} 
%99 (2006); [hep-th/0604025].
%\bibitem{bal}A.P. Balachandran, G. Mangano, A. Pinzul, S. Vaidya, 
%{\it{Int. J. Mod. Phys.}} {\bf{A 21}} 3111 (2006); [hep-th/0508002].
%\bibitem{sun2}B. Chakraborty, S. Gangopadhyay, A.G. Hazra, F.G. Scholtz,
%{\it{J. Phys.}} {\bf{A 39}} 9557 (2006); [hep-th/0601121].
%\bibitem{spal1}A. Smailagic, E. Spallucci, {\it{J. Phys.}} {\bf{A 36}} L467 (2003).
%\bibitem{klaud}J.R. Klauder, B. Skagerstam, {\it{Coherent states :}} {\it{Applications in Physics and Mathematical Physics}} (World Scientific, Singapore, 1985). 
%\bibitem{Majhi1} R.Banerjee, B.R.Majhi and S.Samanta, {\it Phys. Rev.} {\bf D 77}, 124035 (2008) [arXiv:0801.3583]. 
%\bibitem{Paddy}K.Srinivasan and T.Padmanabhan, {\it Phys. Rev.} {\bf D 60}, 024007 (1999) [arxiv:gr-qc/9812028].
%\bibitem{fur}D.V. Fursaev, {\it{Phys. Rev.}} {\bf{D 51}} R5352 (1995); [hep-th/9412161].

  

\end{thebibliography}
\end{document}